\documentclass[prd,aps,nofootinbib,showpacs,showkeys]{revtex4}
\usepackage{graphicx,epsf,amsfonts,amssymb,amsbsy}
\begin{document}

\title{ \bf Quark droplets stability
induced by external magnetic field}
\author{D.~Ebert}
\email{debert@physik.hu-berlin.de}
\affiliation{Institut f\"ur Physik,
Humboldt-Universit\"at zu Berlin, 10115 Berlin, Germany}
\author{K.\,G.~Klimenko}
\email{kklim@mx.ihep.su}
\affiliation{Institute
of High Energy Physics, 142281, Protvino, Moscow Region, Russia}

\begin{abstract}
The influence of a constant homogeneous external magnetic field $H$
on the formation and stability of quark droplets
is investigated within a simple Nambu -- Jona-Lasinio
model by using a thermodynamic approach. For a vanishing magnetic
field stable quark droplets, which are schematically the bags of
massless quarks, are allowed to exist only at $G>G_{bag}$, where $G$
is the quark coupling constant, $G_{bag}=1.37G_{crit}$, and
$G_{crit}$ is the value of the coupling constant above which chiral
symmetry is spontaneously broken down. On the other hand, a
nonvanishing external magnetic field can induce
the stability of
quark droplets so that they may exist even at $G<G_{bag}$. In this
case, depending on the value of $H$, quark droplets are composed
either of massive or massless quarks.
\end{abstract}

\maketitle

\section{ Introduction}
At present time, it is generally accepted that QCD is the
theory of strong interactions between mesons and baryons.
Moreover, it should describe the physical
phenomena in dense hadronic matter as well. Since the
usual (perturbative) methods of QCD are applicable only at
sufficiently high values of the baryon density $\rho$,
in the region of low and moderate densities one should use
nonperturbative methods, e.g. lattice QCD calculations \cite{fodor},
or effective field theory considerations.
The latter ones are especially useful if the influence of external
magnetic field on the properties of dense hadronic matter is
investigated (in this case the incorporation of strong magnetic field
into lattice calculations is not eleborated sufficiently).
It is well-known that at low (baryon) densities
$\rho<\rho_0$, where $\rho_0=0.16$ fm$^{-3}$
is the normal density of baryons inside nuclei,
hadronic matter is just the ordinary nuclear matter.
The widely-known relativistic effective theory for nuclei and
nuclear matter is the Walecka model \cite{wal,kap}.
In this model the protons and neutrons, i.e. the simplest baryons
(nucleons), are point particles. Moreover, the existence of stable
objects with large baryonic number,
which are droplets of simplest massive baryons and hence might be
identified with nuclei, is predicted in the framework of it
with the help of the condensed matter (thermodynamic) approach.

Unfortunately, the Walecka model is unrealistic
in the region of
moderate densities near to the chiral phase transition,
where it is necessary to take
into account the internal quark structure of protons and neutrons.
Similar conditions exist in the
experiments on relativistic ion-ion collisions, where it is expected
that for a short time stable clumps of quark matter (usually called
quark droplets) with rather large baryonic number might be
created. Another example is
presented by neutron stars, where under extreme pressure nucleons
might fuse together, forming droplets of quarks (evidently, the
baryonic density inside quark droplets is larger, than $\rho_0$).

One of the widely-known effective theory of quark-hadronic matter
is the Nambu -- Jona-Lasinio (NJL) model \cite{njl}, which was
successfully used in describing low-energy meson physics
\cite{ebvol}, diquarks \cite{ebka,vog} and meson-baryon interactions
\cite{ebjur,reinh}. In the framework of the NJL model quarks are
treated as elementary point-like particles, whereas
mesons are collective quark-antiquark exitations of its ground
state. This model can also serve as a starting point for
investigations of the mass spectrum of the octet and decuplet of
baryons (see, e.g. \cite{hk,ratti,oettel}).
However, in order to be applicable to dense quark-hadronic
matter, the NJL
model should predict the existence of stable multiquark
droplets as well, i.e. of stable finite-size objects in space, which
are composed of more than 3 quarks and do not tend to shrink
to a point or to expand over the whole space.
One particular way to investigate such objects in the NJL model and
which we will follow in this paper is based on the use of the general
condensed matter approach restricted to vanishing temperature $T$.
In this case a stable multiquark object of the system
is identified with a droplet of the dense quark phase,
surrounded by the vacuum phase with zero
baryonic density. Hence, if there is the thermally
equilibrated coexistence of these two phases,
then one may conclude about the existence of stable quark droplets.
Just in this way, the problem of quark matter
stability was investigated at zero external magnetic field $H$ in
\cite{bub,mark,satarov} in the framework of a chirally symmetric
NJL model. Let us quote the two main results of these
papers: i) at $H=0$, stabilization of (multi)quark droplets is only
possible, when they consist of massless quarks,
ii) the stability requirement puts strong constraints on
the acceptable values of model parameters (coupling constant $G$,
cutoff parameter $\Lambda$ etc). In particular, it was shown that
in the thermodynamic approach based on the NJL model, quark droplets
can be schematically interpreted as MIT-type bags. However, in the
MIT 3-quark bag model the bag constant is a phenomenological input
parameter \cite{chodos}, whereas in the NJL model it is a dynamical
quantity originating from the dynamical breaking of chiral symmetry.

Taking into account the fact that inside neutron stars the magnetic
field strength might reach extremely high values (the influence of
ultra-strong magnetic fields on physical processes inside neutron
stars and on properties of dense quark matter are under permanent
investigations \cite{chak,tatsumi,berg,blashke}), it seems to
us important to reinvestigate the problem of the quark droplet
stability in the framework of NJL-type models at $H\ne 0$. In the
present paper, we study, in particular, the influence of a constant
homogeneous external magnetic field $H$ in a simple NJL model with
two quark flavors. In Section 2, the stability constraints on the
model parameters as well as on the chemical potential $\mu$ are
presented. Apart from this, the phase structure of the model is found
at $T=0$, $H=0$ and $\mu\ne 0$. On this basis, it
is shown that at $H=0$ stability of multiquark droplets, which
consist here only of massless quarks, occurs at $G>G_{bag}\equiv
1.37G_{crit}$ with $G_{crit}=\pi^2/(3N_f\Lambda^2)$
and $N_f$ being the number of flavored quarks. In Sections 3, 4 and 5
the problem of the existence of multiquark droplets
is discussed at $H\ne 0$. In particular, it is argued
on the basis of our earlier papers \cite{k3,k2} that for each value
$G$ smaller than $G_{bag}$, the stability of quark droplets is
induced by an external magnetic field. Moreover, for each $G\in
(G_{crit},G_{bag})$ it is shown that in contrast to the case $H=0$
\cite{bub}, there exists an external magnetic field value
$H_\alpha\sim 10^{19}$ Gauss such, that at $0<H<H_\alpha$
multiquark droplets may be formed from massive quarks. Moreover, we
have shown using energetical arguments that the
magnetic field promotes the creation of stable quark droplets and
enhances the stability of it even at $G>G_{bag}$.
 
\section{Quark droplet stabilization at $H=0$}

\subsection {NJL model and stability constraint}

Let us first give several (very approximative) arguments motivating
the chosen
structure of our QCD-motivated NJL model introduced below.
For this aim, consider two-flavor QCD, $N_f=2$, with nonzero chemical
potential and color group $SU_c(N_c)$ (in our case $N_c=3$).
By integrating in the generating functional of QCD over gluons
and further ``approximating'' the nonperturbative gluon propagator by
a $\delta-$function, one arrives at an effective local chiral
four-quark interaction of the NJL type describing low energy hadron
physics. Finally, by performing a Fierz transformation of the
interaction term and taking into account only scalar and
pseudo-scalar $(\bar q q)$ interactions, one obtains a four-fermionic
model given by the following Lagrangian (in Minkovski space-time
notation) \footnote{The most general Fierz transformed four-fermion
interaction would include additional vector and axial-vector $(\bar q
q)$ as well as scalar and pseudo-scalar, vector and axial-vector-like
$(q q)$-interactions. However, in order to simplify the studying of
the quark matter stability problem in the presence of external
magnetic field, these additional terms are omitted. The role of the
interaction in the diquark channel is discussed in the last Section
5.}
\begin{eqnarray}
 L=\bar q[i\gamma^\nu \partial_\nu+\mu\gamma^0]q
  +G[(\bar qq)^2+(\bar qi\gamma^5\vec \tau q)^2].
  \label{1}
\end{eqnarray}
In (\ref{1})  $\mu >0$ is the
chemical potential of quarks which is the same for all
flavors. Furthermore, we use the notations  $\vec \tau\equiv
(\tau^{1},\tau^{2},\tau^3)$ for Pauli matrices in the flavor space.
Clearly, the model (\ref{1}) is invariant under the continuous chiral
$SU(2)_L\times SU(2)_R$ group. Apart from this, it is invariant
under the discrete chiral transformation, $q\to i\gamma^5 q$.
(Since the baryonic charge is conserved, there exists also a
corresponding $U(1)_V$-invariance of the model). At tree level, the
Lagrangian (\ref{1}) contains besides of the chemical potential only
one free model parameter, i.e. the coupling constant $G$. Clearly,
when including quantum effects (loops) and performing necessary
regularizations, an additional cutoff parameter $\Lambda$ will
appear.

An illustrative way for the investigation of stable quark droplets in
quantum field theory is based on a schematic analogy with the
thermodynamical approach to the liquid-gas first order phase
transition. In such a picture one can consider the quark droplet as a
droplet of the dense liquid phase of the system, whereas the
low-density gas phase corresponds to the vacuum (the zero density
state) which surrounds the quark droplet (strictly
speaking, such an analogy makes really sense only for systems at
finite temperature). Moreover, we suppose
that the two phases are in
the thermal and chemical equilibrium. Of course, in this way, i.e.
when ignoring surface effects of droplets, we have an adequate
consideration of the quark matter stability problem only for quark
droplets which contain a very large number of quarks inside a large
volume (for comparison see \cite{kiriyama}, where surface effects of
droplets were taken into account). It turns out that not for all
values of input parameters of the system this schematic picture of
the existence of stable quark droplets is realized. Thus, the
stability problem of quark droplets arises, which might be expressed
in terms of several quantitative constraints on the parameters of the
system. To solve these constraints means to find the values of the
parameters of the system at which there might appear stable quark
droplets at moderate densities. Let us next describe these stability
constraints.

Recall first that in the thermal and chemical equilibrium between
vacuum and dense quark matter phases their (critical) chemical
potentials, denoted as $\mu_c$, are equal. Then, from the analogy
with condensed matter physics, one can get the following relations
between thermodynamical and dynamical parameters at which the two
phases do coexist (called below stability constraints):
\begin{equation}
m_{dense}<\mu_c<m_{vac}.
\label{stab}
\end{equation}
Here $m_{vac}$ is the dynamical mass of one-quark excitations
of the vacuum and $m_{dense}$ is the mass of quarks in the dense
phase, i.e. inside the quark droplets (see also \cite{bub}).
\footnote{Indeed, one can interpret the chemical potential of any
fermionic system as the lowest energy, needed for the particle in
order to abandon the system. The equilibrium value $\mu_c$ must be
smaller than the mass of the vacuum one-particle exitations
$m_{vac}$ (such particles have enough energy for leaving the vacuum,
which is empty due to this reason). However, in order that quarks do
not leave the dense phase, their mass $m_{dense}$ must be smaller
than $\mu_c$ -- the minimum particle energy, necessary for leaving
the system.}  Note that the values of $m_{dense}, \mu_c, m_{vac}$ are
supplied by the thermodynamic potential $\Omega (m;\mu)$ of the
system, where the average of the composite scalar $\sigma$-field
$m=<\sigma>$ is the order parameter of spontaneous chiral symmetry
breaking which in the thermal equilibrium is the global
minimum point $m_o$ of the function $\Omega (m;\mu)$ versus $m$.
Clearly, $m_o$ equals to the dynamical quark mass
which is determined by the gap equation $m_o=-2G<\bar qq>$.
(Obviously, if $m_o=0$, the ground state is chirally invariant.)
Hence, in order to find all the admissable values of parameters of
the model, at which stable quark droplets could exist in the quantum
system at moderate density, it is sufficient to solve the stability
constraint (\ref{stab}) with respect to these parameters. On this
route one should perform several technical steps: i) get the
thermodynamic potential $\Omega (m;\mu)$; ii) find the global
minimum point of $\Omega (m;\mu)$ vs $m$ and establish
its dependence on the external parameters of the theory thereby
obtaining the phase structure of the model; iii) choose all the
first order phase transition critical points $\mu_c$ obeing the
constraint (\ref{stab}). \footnote{At the critical line $\mu_c$ of
the 1st order phase transition, the order parameter $m_o$ changes its
value by a jump, i.e. at this line it is a discontinuous function vs
input parameters of the model. At the critical line of the
second order phase transition the order parameter is a continuous
function, but first- or higher order derivatives of it are
discontinuous.} The corresponding values of allowed parameters
obtained in this way are just the physically acceptable ones, at
which stable quark matter might be described by the initial model.

Now let us carry out this scheme on the basis of the chosen NJL
model.

\subsection{Solution of the stability constraint at $H=0$}

In the mean field approximation, the thermodynamic potential of the
 model (1) is given by the expression \cite{bub,k2,zhuang}:
\begin{eqnarray}
\Omega(m;\mu)=\frac{m^2}{4G}-
2N_cN_f\int\frac{d^3p}{(2\pi)^3}\biggl\{E_p+
\theta(\mu-E_p)(\mu-E_p)\biggr\},
 \label{8}
\end{eqnarray}
where $E_p=\sqrt{m^2+\vec p^{~2}}$, $N_c$ is the number of colors,
i.e. $N_c=3$, and $N_f=2$. Using for the divergent integral in
(\ref{8}) a Lorentz noninvariant regularization (this procedure is
simply the replacement of the infinite integration region in
(\ref{8}) in favour of the compact one: $\vec p^{~2}\leq \Lambda^2$,
where the cutoff parameter $\Lambda$ is an additional parameter of
the theory to be fitted by the experimental data),
one can find the following expression
\begin{eqnarray}
&&\Omega(m;\mu)=\frac{m^2}{4G}
-\frac{3N_f}{8\pi^2}\left
[\Lambda(2\Lambda^2+m^2)\sqrt{m^2+\Lambda^2}
-m^4\ln\left (\frac {\Lambda+\sqrt{m^2+\Lambda^2}}{m}\right
)\right ]- \nonumber \\
&-&\frac {N_f\theta(\mu-m)}{8\pi^2}\left
[\mu(2\mu^2-5m^2)\sqrt{\mu^2-m^2}
+3m^4\ln\left (\frac
{\mu+\sqrt{\mu^2-m^2}}{m}\right )\right ].
\label{9}
\end{eqnarray}

\subsubsection{The vacuum case $\mu= 0$.}

In this case, from (\ref{9}) one can get the following stationarity
equation:
\begin{equation}
\frac\partial{\partial m}\Omega(m;0)=\frac{3N_fm}{2\pi^2}
\left\{\frac{\pi^2}{3N_fG}-F(m,\Lambda)\right\}=0,
\label{13}
\end{equation}
where
\begin{eqnarray}
F(m,\Lambda)=\Lambda\sqrt{m^2+\Lambda^2}
-m^2\ln\left (\frac {\Lambda+\sqrt{m^2+\Lambda^2}}{m}\right ).
\label{11}
\end{eqnarray}
The function $F(m,\Lambda)$ monotonically decreases on the interval
$m\in (0,\infty)$ such that $F(0,\Lambda)=\Lambda^2$ and
$F(\infty,\Lambda)=0$ ($F(m,\Lambda)=2\Lambda^3/(3m)+o(\Lambda/m)$
at $m\to\infty$). Hence, equation (\ref{13}) has a nontrivial
solution $m_o(G,\Lambda) \equiv M\neq 0$ if and only if the
constraint $\frac{\pi^2}{3N_fG}-F(0,\Lambda)<0$ (or $G\geq
G_{crit}\equiv\frac{\pi^2}{3N_f\Lambda^2}$) is fulfilled. In this
case, the global minimum of the function $\Omega(m;0)$ lies at $M$.
Thus, both the continuous and discrete chiral symmetries of the
NJL--model are spontaneously broken. (At $G\leq G_{crit}$ the
stationarity equation (\ref{13}) has only the trivial solution $m
=0$. Thus, in this case the vacuum is chirally invariant, and
fermions have zero mass. Since this phase is not adequate to the real
QCD vacuum at $T=\mu=0$, we will not consider in detail
the undercritical values of the coupling constant in the present
paper.) Evidently, among the three parameters $\Lambda, G, M$ only
two are independent ones at $G\geq G_{crit}$. In the following it is
sometimes convenient to use $\Lambda$ and $M$ as
independent free parameters of the model at $G\geq G_{crit}$.
Clearly, $M$ is here the mass of the one-quark excitation
of the chirally-noninvariant vacuum with zero baryon density which,
at $H=0$, is just the quantity $m_{vac}$ from (\ref{stab}). Then,
$G=\frac{\pi^2}{3N_f F(M,\Lambda)}$ and there is a one-to-one
correspondence between $G>G_{crit}$ and $M>0$.

\subsubsection{The case $\mu\ne 0$.}

Note first of all that for $G\leq G_{crit}$ and arbitrary values of
$\mu$, the chiral symmetry is not spontaneously broken down.
However, above the critical coupling constant the situation is more
complicated. Recall, that in this case the phase structure of the
simplest NJL model with $N_f=1$ and Lorentz invariant regularization
was investigated in \cite{k1}. Using the same
methods, it is easy to obtain the phase structure of model (\ref{1})
for the Lorentz-noninvariant regularization considered above.
Thus, without any preliminary calculational discussions, we present
the phase portrait of the NJL model (1) schematically on Fig.
1. On this figure one can see the $(M,\mu)$-plane, which is divided
into three domains: A, B, C. There is a symmetric dense phase with
massless quarks for points $(M,\mu)\in$ A. In the region
B, the phase with zero baryon density is realized. In this phase the
chiral symmetry is spontaneously broken down, and quarks aquire a
mass $M$. (In this sense, phase B models the real QCD vacuum.)
Finally, the region C on this figure corresponds to the dense
chirally noninvariant phase, in which quarks have a nonzero
$\mu$-dependent mass. The boundaries between these phases are the
critical lines of second order phase transitions (solid lines on Fig.
1), or first order ones (dashed curves on Fig. 1).
In particular, the curve $l_{BC}=\{(M,\mu):\mu=M\}$ is the critical
line of the second order phase transitions, since the second order
derivative $\partial^2 m_o/(\partial\mu)^2$ of the global minimum
point $m_o$ of the potential $\Omega (m;\mu)$ is a discontinuous
function vs $\mu$ on this curve.
The curve $l_{AB}=\{(M,\mu):\mu=\mu_c(M)\}$ (all points of this
critical curve lie below the line $\mu=M$) is defined by the equation
\begin{equation}
\Omega (M;\mu)=\Omega (0;\mu),
\label{16}
\end{equation}
i.e. on this line the order parameter of the system -- the global
minimum point $m_o$ -- changes its value by a jump from $M$ to 0, and
a first order phase transition takes place. Solving (\ref{16}) in the
region $\mu\le M$, one can obtain the manifest form of this curve:
 \begin{equation}
 \mu_c(M)=\left\{3\Lambda^3\sqrt{M^2+\Lambda^2}-3\Lambda^4-\frac32
 M^2
 F(M,\Lambda)
 \right\}^{1/4},~~~\mbox{where}~~~M\ge M_c=0.56\Lambda .
 \label{17}
 \end{equation}
Moreover, there are two tricritical points $\alpha$ and
$\beta$ on this phase portrait. \footnote{A point of the phase
diagram is called a tricritical one if, in an arbitrary small
vicinity of it, there are first- as well as second-order phase
transitions.} Their coordinates are:
$\alpha=(\tilde M,\tilde\mu)$, $\beta=(M_c,M_c)$, where
$\tilde M=0.31\Lambda$, $\tilde\mu=0.37\Lambda$ and
$M_c$ is defined in (\ref{17}).

Now one can easily see that constraints (\ref{stab}) are fulfilled
at $M\ge M_c$ (or in terms of coupling
constants $G$ and $G_{crit}$, stability constraints (\ref{stab})
are valid only at $G>G_{bag}\equiv 1.37G_{crit}$). Indeed, for each
such value of $M$ there is a corresponding critical value of the
chemical potential $\mu_c(M)$ (\ref{17}) (recall, that
$0<\mu_c(M)<M$) at which the two phases do coexist. In other words,
in this case stable droplets of the dense phase A are
surrounded by the chirally noninvariant vacuum (phase B) with zero
baryon density. Obviously, the inequalities (\ref{stab}) are valid,
since in the case under consideration $m_{dense}=0$ and $m_{vac}=M$.
\footnote {It might appear that points of the 1st order phase
transitions critical line,  which connects two tricritical points
$\alpha$ and $\beta$ of Fig. 1, are also suitable ones. However, in
this case there is the coexistence of two dense phases, and the
physical vacuum cannot be identified with one of them. So, we reject
the corresponding values of $M\in (\tilde M,M_c)$ at $H=0$.} The
baryon density inside the quark droplet is easily calculated from the
relation:
\begin{equation}
\rho_{drop}(M)=-\frac{\partial\Omega(0;\mu)}{N_c\partial\mu}\vrule
\begin{array}{cc} ~~\\ \scriptstyle{\mu=\mu_c(M)}
 \end{array}=\frac{N_f}{3\pi^2}\mu^3_c(M).
\label{18}
\end{equation}
Now, in the framework of the NJL model, one can
imagine for $M\ge M_c$ the following quark
matter transition in dependence on the average baryonic
density $\rho$: At sufficiently small $\rho$ a rather small amount
of quark droplets, having a rather small size, appears in the space.
The droplets are composed of massless quarks and surrounded by the
vacuum with zero baryon density; however, inside each droplet the
baryonic density  equals to $\rho_{drop}(M)$ (\ref{18}). With growth
of $\rho$ the number of droplets as well as their sizes are
increasing. When $\rho$ reaches the value $\rho_{drop}(M)$, the total
space is occupied by massless quarks.

Finally, notice that the existence of stable quark droplets
surrounded by a chirally noninvariant vacuum was also considered in
the framework of a NJL model in \cite{bub}. However, there the
problem was investigated only for three physically interesting sets
of parameters: $M=M_1\equiv 0.48\Lambda$, $\Lambda =650$ MeV (set 1),
$M=M_2\equiv 0.67\Lambda$, $\Lambda =600$ MeV (set 2) and
$M=M_3\equiv 0.88\Lambda$, $\Lambda =570$ MeV (set 3). Of course, the
corresponding results are reproduced by our ones: for set 1 stable
quark droplets are forbidden, but for sets 2,3 stable droplets are
formed from massless quarks. In particular, it follows from
(\ref{18}) that $\rho_{drop}=2.86\rho_o$ for set 2 and
$\rho_{drop}=4.19\rho_o$ for set 3 (here $\rho_o$ is the ordinary
nuclear matter density). In the present paper we have generalized
these considerations and found the exact interval $M\ge 0.56 \Lambda
\equiv M_c$ corresponding to stable quark droplet formation.

\section{Phase structure at $H\not =0$}

Let us next study the phase structure of the 
NJL model (\ref{1}) containing  $u$- and
$d$-quarks with electric charges $e_1=2|e|/3$ and $e_2=-|e|/3$ ($e$
is the electric charge of electrons), respectively. Due to the
difference in quark electric charges, the continuous chiral symmetry
is violated in the presence of an external magnetic field, but the
discrete one, $q\to i\gamma^5 q$, is not. It is useful to consider
first the vacuum case with $\mu =0$, where the thermodynamic
potential of the system looks like:
\begin{equation}
\Omega(m;0,H)=\frac{m^2}{4G}+\sum_{i=1}^2\frac{N_c|e_iH|}{8\pi^2}
\int_{0}^{\infty} \frac{ds}{s^2} \exp (-sm^2)~\coth(|e_iH|s),
\label{29}
\end{equation}
where $N_c=3$.
This expression can easily be transformed into the following one
 (see, e.g., \cite{k2}):
\begin{equation}
\label{30}
\Omega(m;0,H)=\Omega
(m;0)-\sum_{i=1}^2\frac{3(e_iH)^2}{2\pi^2}\Bigl\{\zeta
'(-1,x_i)-\frac 12[x_i^2-x_i]\ln x_i
+\frac{x_i^2}4\Bigr\},
\end{equation}
where $\Omega (m;0)$ is given in (\ref{9}) at $\mu=0$ and
$x_i=m^2/(2|e_iH|)$. Moreover, $\zeta'(-1,x)$$=d\zeta(\nu,x)
/d\nu|_{\nu=-1}$, where $\zeta (\nu,x)$ is the generalized Riemann
zeta-function \cite{bateman}. The thermodynamic potential
(\ref{29})-(\ref{30}) is an even function with respect to $m$. This
fact reflects the symmetry of the original model under discrete
chiral transformation $q\to i\gamma^5 q$ in the presence of an
external magnetic field. Obviously, this symmetry might be
spontaneously broken if the global minimum point $m_o$ of $\Omega
(m;0,H)$ is unequal to zero, but remains intact if $m_o=0$. In the
case $G>G_{crit}$, the global minimum point of $\Omega (m;0,H)$ is
the solution of the stationarity equation (see \cite{k2}):
\begin{equation}
\label{31}
\frac{2\pi^2}{3}\frac {\partial}{\partial m}\Omega (m;0,H)
=m\{2F(M,\Lambda)-2F(m,\Lambda)-I_1(m)-I_2(m)\}=0,
\end{equation}
where the function
$F(x,\Lambda)$ is defined in (\ref{11}),
\begin{eqnarray}
\label{32}
I_i(m)=|e_iH|\{\ln\Gamma(x_i)-\frac 12\ln(2\pi)+x_i-\frac
12(2x_i-1)\ln
x_i\}
\end{eqnarray}
and $\Gamma(x)$ is the first order Euler gamma-function
\cite{bateman}.
Below we suppose that $H$ is a nonnegative quantity.

The function $F(x,\Lambda)$ vs $x$ monotonically decreases on
the interval $x\in (0,\infty)$ (see subsection II.B), and the
functions $I_i(x)$ vs $x$ also monotonically decreases on the same
interval from $+\infty$ to 0 (see \cite{k2}). Hence, for arbitrary
fixed values of $H,M$ there exists only one nontrivial solution
$m_o(H)$ of equation (\ref{31}), which is the global minimum point of
$\Omega (m;0,H)$. The quantity $m_o(H)$ is a monotonically
increasing function of $H$ and $m_o(H)\to M$ at $H\to 0$.
It means, that at $G>G_{crit}$ (or for all $M>0$) the external
magnetic field $H$ enhances the discrete chiral symmetry breaking.
Note that in the presence of an external
magnetic field the parameter $m_{vac}$ from (\ref{stab}) is equal to
$m_o(H)$, i.e. depends on the values of $H$. \footnote{At
$G<G_{crit}$, the magnetic catalysis effect of dynamical chiral
symmetry breaking (DCSB) takes place.
This means that at $H=0$ the NJL vacuum is chirally symmetric, but an
arbitrary small value of the external magnetic field induces
DCSB, and fermions acquire a nonzero mass $m_o(H)$ \cite{mir,k3}
(see also the reviews \cite{gus} and  recent papers on
applications of the magnetic catalysis effect \cite{incera}).}

Now let us consider the more general case, when $H\neq 0,\mu\not =
0$. Using the methods of \cite{k2}, it is possible to obtain
the thermodynamic potential:
\begin{equation}
\label{33}
\Omega (m;\mu,H)=\Omega
(m;0,H)-\sum_{i=1}^2\frac{3|e_i|H}{4\pi^2}\sum_{k=0}^{\infty}
\alpha_k
\theta (\mu-s_{ik})\Biggl\{\mu\sqrt{\mu^2-s_{ik}^2}-s_{ik}^2\ln\left
[\frac {\mu+\sqrt{\mu^2-s_{ik}^2}}{s_{ik}}\right ]\Biggr\},
\end{equation}
where $s_{ik}=\sqrt{m^2+2|e_i|Hk}$, $\alpha_k=2-\delta_{0k}$,
and $\Omega (m;0,H)$ is given in (\ref{30}). The
corresponding stationarity equation at $G>G_{crit}$ looks like:
\begin{eqnarray}
\frac {2\pi^2}{3}\frac {\partial}{\partial m}\Omega(m;\mu,H)&\equiv&
m\Biggl \{2F(M,\Lambda)-2F(m,\Lambda)-I_1(m)-I_2(m)+
\nonumber \\
&+&\sum_{i=1}^2|e_i|H\sum_{k=0}^{\infty} \alpha_k \theta
(\mu-s_{ik})\ln\left
[\frac {\mu+\sqrt{\mu^2-s_{ik}^2}}{s_{ik}}\right ]\Biggr\}=0.
\label{34}
\end{eqnarray}
It is useful to divide the $(H,\mu)$-plane into regions $\omega_k$,
which have the following form:
\begin{equation}
\label{35}
(H,\mu)=\bigcup\limits^{\infty}_{k=0}\omega_k,~\,\,\mbox{where}~~~
\omega_k=\{(H,\mu):2|e_1|Hk\leq \mu^2\leq 2|e_1|H (k+1)\}.
\end{equation}
Then, the construction of the phase portrait is greatly
simplified, since for each region $\omega_n$ from (\ref{35}) only the
terms with $k=0,..,n$ from the $e_1$-sums in (\ref{33})-(\ref{34})
are nonvanishing, and only the terms corresponding to $k=0,..,2n+1$
from the $e_2$-sums in these expressions are nonzero as well.
Hence, it is possible to perform
step-by-step the numerical investigation of the phase structure of
the NJL model in the regions $\omega_0$, $\omega_1$, etc.
In particular, we have considered in detail the phase structure
for the same three values of model parameter $M=M_1,M_2,M_3$ as in
the paper \cite{bub}.

In the case $M=M_1$, we have studied the phase structure
only in the region $\omega_0$ (\ref{35}). The phase portrait in
terms of $(H,\mu)\in\omega_0$ is presented schematically on Fig.
2, where $L=\{~(H,\mu):~\mu=m_o(H)\}$ denotes the upper estimate of
the value $\mu_c$ from (\ref{stab}), i.e. $m_{vac}=m_o(H)$. Below
$L$ the $(H,\mu)$-plane is divided into four domains A, B, C and
C$_1$. The solid and dashed lines on this figure correspond to the
critical curves of second and first order phase transitions,
respectively. The points from the region A correspond to the
massless, chirally invariant and dense phase. In region B there is
the chirally noninvariant phase with zero baryon density. The
one-quark excitations of the vacuum phase B have a mass equal to
$m_o(H)$, which depends on $H$ and has no $\mu$-dependency (the phase
B can be identified with the real QCD vacuum without baryons).
On Fig. 2 one can see also two another massive, i.e. chirally
noninvariant, dense phases C and C$_1$ as well as three tricritical
points $\alpha,\beta,\gamma$. Their coordinates are
$\sqrt{2|e|H_\alpha} /\Lambda=0.85$, $\mu_\alpha /\Lambda=0.45$,
$\sqrt{2|e|H_\beta} /\Lambda=0.82$, $\mu_\beta /\Lambda=$$0.47$ and
$\sqrt{2|e|H_\gamma} /\Lambda=0.67$, $\mu_\gamma /\Lambda=0.48$.

Now let us draw special attention to the
boundaries $l_{AB}$ and $l_{BC}$ of the vacuum phase B. The boundary
$l_{AB}$ between phases A and B is the graph of the function
$\mu=\mu_c(H)$, where $\mu_c(H)$ is the solution of the equation
\begin{equation}
\Omega(0;\mu,H)=\Omega(m_o(H);\mu,H),
\label{26}
\end{equation}
i.e. on this curve the order parameter -- the global minimum point of
the thermodynamic potential -- changes by a jump its
value (from $m_o(H)$ to 0), and a first order phase transition
takes place. It turns out that the critical line $l_{AB}$ as a whole
is arranged inside the $\omega_0$-region. However, the boundary
$l_{BC}$, which is also a critical line of first order phase
transitions, where the order parameter changes by a jump
from $m_o(H)$ to some another definite nonzero value $m_1(H)$, is
stretched through the regions $\omega_0$,.., $\omega_{100}$ at least.
Evidently, it tends to the point $M_1$ (see Fig. 2), but to prove
this hypothesis, we would need more calculational capacities. The
curve $l_{BC}$ is the graph of the function $\mu=\tilde\mu_c(H)$,
where $\tilde\mu_c(H)$ is defined by the equation
\begin{equation}
\Omega(m_1(H);\mu,H)=\Omega(m_o(H);\mu,H).
\label{261}
\end{equation}

At $M=M_{2,3}$ and $H\ne 0$ the phase portraits at
$\mu<m_o(H)$ is qualitatively the same as in the case $H=0$:
there are two phases A and B in this region, and the boundary
$l_{AB}$ is the curve of first order phase transitions.

\section{Quark droplet stability at $H\ne 0$}

In the previous section the phase structure of the NJL model
was investigated at $H\ne 0$ and for three particular values of the
model parameter $M$: $M=M_1\equiv 0.48\Lambda$, $M=M_2\equiv 0.67
\Lambda$ and $M=M_3\equiv 0.88 \Lambda$. Now, putting special
attention just to these values of $M$, let us discuss the quark
matter stability problem in the presence of an external magnetic
field. Stability means that for fixed $H$ there is the critical value
$\mu_c$ of the chemical potential at which the vacuum and dense quark
phases coexist, i.e. at $\mu_c$ a first order phase transition
occurs. In this case, the value $\mu_c$ must satisfy the stability
relations (\ref{stab}).

\subsection{Magnetic catalysis of quark matter stability}

First of all, let us draw our attention to the case $M=M_1$.
The corresponding phase portrait is shown on Fig. 2,
where points $(H, \mu=m_o(H))$ are located on the curve L, and
$m_o(H)$ is the global minimum point of $\Omega$ at $\mu=0$. Hence,
$m_o(H)$ is the mass of the one-quark exitations of the empty
vacuum phase B and must be identified with $m_{vac}$ from
(\ref{stab}). The boundary of this phase is the first order phase
transition line which consists of two curves $l_{AB}$ and $l_{BC}$
and lies below the line L.
\footnote{On Fig. 2 there is another  critical line,
where 1st order phase transitions take place as well. It is the line
which connects tricritical points $\beta$ and $\gamma$.
 However, it is the boundary between two dense phases, and no
one of them might be identified with the physical vacuum.} So, at $H>
H_\alpha$, there is coexistence of the zero density
vacuum phase B with the dense phase A  on the curve $l_{AB}$ (the
one-quark exitations of the phase A have zero mass). Now, if one puts
$m_{dense}\equiv 0$, then $\mu_c \equiv \mu_c(H)$
satisfies the stability constraints (\ref{stab}), where $\mu_c(H)$ is
defined by the equation (\ref{26}). Hence, at $H>H_\alpha$,
multiquark droplets which are schematically the bags of massless
quarks, are allowed to exist in the vacuum.

At $H<H_\alpha$ and in the case $(H,\mu=\tilde\mu_c(H))\in l_{BC}$,
where $\tilde\mu_c(H)$ is defined by the equation (\ref{261}),
the function $\Omega(m;\mu,H)$ has two different global minimum
points: the first one is located at $m=m_o(H)$, the second one at
some definite nonzero point $m_1(H)$. So, at $H<H_\alpha$ there is a
coexistence of the vacuum phase B with the dense phase C  on the
curve $l_{BC}$ (the value $m_1(H)$ is just the mass of the one-quark
exitations of the ground state of the phase C). Now, if one puts
$m_{dense}\equiv m_1(H)$, $m_{vac}\equiv m_o(H)$, then $\mu_c\equiv
\tilde\mu_c(H)$ satisfies the stability constraints (\ref{stab}).
This conclusion can be obtained from a detailed analysis of the
stationarity equation (\ref{34}). At $\Lambda =650$ MeV it is also
confirmed by numerical calculations depicted on Fig. 3 showing
graphs of the functions $m_{vac}$, $\mu_c$ and $m_{dense}$ vs $H$
corresponding at $\sqrt{2|e|H}/ \Lambda <0.85$ to those points of the
line $l_{BC}$, which lie inside the region $\omega_0$. So, at
$H<H_\alpha$ stable quark droplets composed of {\it massive}
quarks are allowed to exist in the NJL model.

Recall, that in the case $M=M_1$ the stability of quark
droplets is absent at $H=0$
(see Section II as well as \cite{bub}). However, as we have
proved, 
in the presence of an external magnetic field there might exist
stable quark droplets, i.e. an external
magnetic field promotes the stabilization of quark matter. Moreover,
it follows from our consideration that at
$H\ne 0$, in contrast to the case $H=0$, stable multiquark
clumps might be composed both from massless quarks (at $H>H_\alpha$),
and {\it massive} ones (at $H<H_\alpha$).  These are the main results
of our paper.

Remark, that usually the value of the cutoff parameter $\Lambda$
varies in the interval $(0.5\div 1)$ GeV. In this case,
the magnitude of $H_{\alpha}$ can be estimated to be
of the order $10^{19}$ Gauss. \footnote{This estimate can be obtained
from the known relation $m_e^2/|e|=4.4\times 10^{13}$ Gauss (where
$m_e$ and $e$ are the mass and the electric charge of an electron,
respectively), which permits to connect (MeV)$^2$ and Gauss
units.} According to modern estimates, the strength
of the magnetic field at the surface of neutron stars can reach
$10^{15}$ Gauss, and at the star core it may be as large as
$10^{18}$ Gauss \cite{duncan}. Hence, for $M=M_1$ as well as for some
neighbouring values of $M$, the NJL model may predict
the existence of stable multiquark clumps, composed of
{\it massive} quarks, inside neutron stars.

Note finally, that for arbitrary values of $M\in (\tilde M,M_c)$
(see Fig. 1), quark droplet stabilization is induced by the external
magnetic field in the same manner, as in the case $M=M_1$.

At $M=M_{2,3}$ and in the presence of an external magnetic field, the
zero density vacuum phase B ($m_{vac}=m_o(H)$) might coexist only
with the dense phase A ($m_{dense}=0$) at $\mu =\mu_c$. Critical
values $\mu_c$ of the chemical potential in these cases are depicted
on Figs 4, 5 for $M=M_{2}$ and $M=M_{3}$, respectively. It follows
from these figures that the stability constraints (\ref{stab}) are
also fulfilled, so the stable quark droplets are allowed to exist. As
in the case $H=0$ (see \cite{bub}), at $M=M_{2,3}$ and $H\ne 0$ the
finite lumps of stable quark matter are  schematically the bags of
massless quarks only.

 The phase structure of the NJL model at $H\ne 0$ and $\mu\ne 0$
was studied in \cite{k3,k2}, but for qualitatively other values of
model parameters. In particular, the  stability problem of multiquark
clumps was not considered in these articles. Thus, in view
of our present results, let us now discuss it in these cases, too.
In \cite{k3}, in connection with the magnetic catalysis effect, the
influence of an external magnetic field on the dense NJL model was
investigated at $G<G_{crit}$. It turns out that at sufficiently small
values of $\mu$ and arbitrary values of $H$ the external magnetic
field induces the rearrangement of the model vacuum: at $H=0$ it is
chirally invariant, but the chiral symmetry breaking is induced at
$H\ne 0$ (magnetic catalysis effect \cite{mir,gus}).
Due to the zero baryonic density it is the vacuum phase B.
At larger values
of $\mu$ the dense chirally symmetric phase A with massless quarks
is realized. As shown in \cite{k3}, a first order phase
transition occurs between the A and B phases. Now one can easily
check that the stability condition (\ref{stab}) of the present paper
is fulfilled at $G<G_{crit}$, $H\ne 0$, too. Hence, in this case
stable droplets, composed from massless quarks,
are allowed to exist. These droplets are surrounded by the
chirally noninvariant vacuum.
In papers \cite{k2} the parameter $M$
was fixed in such a way, that $M<\tilde M<M_c$ (see Fig. 1 of the
present paper), i.e. at $H=0$ the quark matter is not stable in
this case. Then, a rather complicated
phase portrait was shown to exist in the $(H,\mu)$-plane. A more
detailed analysis results in the statement that the stability
constraints (\ref{stab}) can be solved for these values of model
parameters as well (in a similar way, as in the case $M=M_1$).

Hence, at $H\ne 0$, in contrast to the $H= 0$ case, stable quark
droplets are allowed to exist in the framework of NJL model (1) for
all values of the coupling constant $G>0$: at $G<G_{bag}$ this
property is induced by an external magnetic field, whereas at
$G>G_{bag}$ it is an own property of the model.

\subsection{Energetical approach to the stability problem}

It might seem that at $M=M_{2,3}$ and $H\ne 0$ the solution of the
quark matter stability problem is qualitatively the same as at $H=0$.
However, there are some peculiarities which can be found
by an energetical consideration of this problem.
Analogously to the case at $H=0$ \cite{bub}, let us define
the energy density  $\varepsilon$ at $H\ne 0$ by the following
relation
\begin{eqnarray}
\varepsilon = \Omega (m;\mu,H)-\Omega (m_o(H);0,H)+N_c\mu\rho,
\end{eqnarray}
where $\rho=-\partial\Omega /(N_c\partial\mu)$ is the baryonic
density and $m_o(H)$ -- the dynamical quark mass in the vacuum phase
B. A very important physical quantity is $\varepsilon/\rho$ -- the
energy per one baryon. At $H=0$ the graph of this function vs $\rho$
is presented on Fig. 6 for $M=M_2$ and $\Lambda =600$ MeV (see
also \cite{bub}). The absolute minimum of this function lies at the
point $\rho =\rho_{drop}(M_2) \equiv 2.86\rho_0$ which is equal to
the baryonic density inside stable quark droplets. At the point
$\rho=0$ there is a local minimum of $\varepsilon/\rho$,
corresponding to the real empty physical QCD-vacuum, i.e. to the
phase B. Due to the configuration of the curve $\varepsilon/\rho$ vs
$\rho$ at $H=0$, one can conclude that the vacuum, i.e. the state
with $\rho=0$, is a metastable one. It means that not for all values
of baryon density fluctuations the system will pass automatically to
the mixed phase with stable quark droplet formation. Indeed, the
curve $\varepsilon/ \rho$ has a local maximum at some point
$\rho_{max}$. So, if a quark droplet with $\rho<\rho_{max}$ is
created in the vacuum, it is energetically favorable for the
system to reach zero density, i.e. to let tend the 
volume of the quark droplet
to infinity thus pushing  out all quarks from the system. As a
result, one gets the initial empty vacuum. However, if a quark
droplet with $\rho> \rho_{max}$ is formed, then it will be a stable
one, since in this case the droplet simply changes its volume in
order to reach the most energetically favorable density
$\rho_{drop}(M_2)$.

The situation is essentially different in the presence of an
external magnetic field. In this case the graph
of the function $\varepsilon/ \rho$ has a local maximum in the point
$\rho=0$ (see Fig. 6), so at arbitrary small baryonic density
fluctuations the stable quark droplets will appear in the system.
A similar property of the function $\varepsilon/ \rho$ is observed
for the case $M=M_1$ (see Fig. 7) as well as for other values of
the parameter $M$. Hence, the external
magnetic field promotes the creation of the stable quark matter.

Let us denote the global minimum point of the function $\varepsilon/
\rho$ as $\rho_{drop}$ (this value is the baryonic density inside the
stable quark droplets). Then, the binding energy per quark
$E_B$ inside stable droplets is given by the relation
\begin{eqnarray}
E_B\equiv -\left (\left.\frac{\varepsilon}{N_c\rho}\right
|_{\rho=\rho_{drop}}-m_{vac}\right ) =m_{vac}-\mu_c,
\end{eqnarray}
where $m_{vac}=m_o(H)$. Of course, both quantities $E_B$ and
$\rho_{drop}$ depend on $H$. Their graphs are presented on Fig.
\ref{ensv} and Fig. \ref{dens}, respectively. Recall, that in the
case $M=M_2$ the quark binding energy $E_B$ is about $20$ MeV at
$H=0$ (see \cite{bub}). Now one can easily see from Fig. \ref{ensv}
that at nonzero $H$ this quantity might be several times larger, than
at $H=0$. So, the external magnetic field enhances the stability of
quark matter.

Note that on Fig. \ref{dens} the density curve corresponding to the
case $M=M_2$ has a cusp at the point $\sqrt{2|e|H}/ \Lambda \approx
1$. The matter is that in the chirally symmetric phase such
quantities, as particle density, magnetization etc, have a
discontinuity of their derivatives over $\mu$ and $H$, when a new
Landau level is switched on \cite{k2}. In the case under
consideration in addition to the lowest Landau level the first one
for the $d$-quarks is brought into account just at the point
$\sqrt{2|e|H}/ \Lambda \approx 1$, so the cusp for the density curve
appears. (On  Fig. \ref{ensv}, a cusp on the curve $E_B$ vs $H$ also
appears at the point $\sqrt{2|e|H}/ \Lambda \approx 1$  in the case
$M=M_2$, which is, however not so sharp. One can easily see a cusp on
the curve $\mu_c$ vs $H$ of Fig. 4 as well.) Of course, the smaller
the lowest value of the $H$-interval which is taken into account, the
more cusps will appear on the density curve as well as on the graphs
of other physical quantities. At $H\to 0$ there are infinitely many
cusps on these curves, which results in the experimentally observed
oscillating phenomena of corresponding physical quantities (see
\cite{k2}).

\section{Summary and discussions}

In the present paper the role of an external magnetic field $H$ on
the formation of stable quark matter droplets inside the chirally
noninvariant vacuum is investigated in the framework of the simple
NJL model (\ref{1}). Earlier, the problem of quark matter
stability was considered in terms of more general effective theories
with four-fermionic interactions as well, but only at $H=0$
\cite{bub,mark,satarov}. Thus, we began our study with the case of a
vanishing external magnetic field, too. Moreover, we used the same
thermodynamic approach for treating multiquark droplets as in
\cite{bub}. In this case, any multiquark clump is imagined as a
droplet of the dense quark phase of the system, which is surrounded
by a chirally noninvariant vacuum phase, and there is thermal 
and chemical equilibrium between both phases.
This qualitative picture can be transformed into a quantitative
constraints (\ref{stab}) on dynamical quark masses and the chemical
potential at which the coexistence of two phases is possible.
Moreover, it was shown in \cite{bub} that at some
values of the physically acceptable model parameters (coupling
constants, cutoff parameter, etc) the stability constraints
(\ref{stab}) are not fulfilled for the NJL model at $H=0$. In those
cases stable multiquark droplets are forbidden, and the NJL model is
not applicable for the description of dense quark matter.

In our paper we have in the first step obtained
the general solution of the stability constraints (\ref{stab}) at
$H=0$ for the NJL model considered above. Here, we have considered
only the physically interesting case,
when $G>G_{crit}=\pi^2/(3N_f\Lambda^2)$, in which case chiral
symmetry is spontaneously broken down and quarks acquire a dynamical
mass $M$ (at $\mu =0$). Due to the stationarity equation (\ref{13}),
the quantities $M$ and $G$ (at $G>G_{crit}$) are in a one-to-one
correspondence. Thus, we find it convenient to use the value $M$
(instead of $G$) as a model parameter (if $M$ changes from 0 to
$\infty$, then $G$ monotonically increases from $G_{crit}$ to
$\infty$).  Stable quark matter droplets inside the chirally
noninvariant vacuum are then allowed in the model under consideration
only at $M>M_c\equiv 0.56\Lambda$, i.e. at $G>G_{bag}\equiv
1.37G_{crit}$, and they are composed of massless quarks. Notice also
that in \cite{bub} the stability problem was investigated only at
three physically admissable sets of parameters: $M=M_1\equiv
0.48\Lambda$, $\Lambda =650$ MeV (set 1), $M=M_2\equiv 0.67\Lambda$,
$\Lambda =600$ MeV (set 2) and $M=M_3\equiv 0.88\Lambda$, $\Lambda
=570$ MeV (set 3). There it was pointed out that for set 1 stable
quark droplets are forbidden, but for sets 2 and 3 stable quark
matter is allowed to exist. Of course, these results coincide with
our ones at $H=0$.

Next, we took into account an external magnetic field and studied
the problem of quark matter stability in detail at the same
values of the parameter $M$ as in ref.\cite{bub}. It turns out that
at any value of $H$ in the cases $M=M_{2,3}$ stable quark droplets
are allowed to exist. Moreover, as at $H=0$, these multiquark
droplets consist of massless quarks. \footnote{It is necessary to
note, that for such values of $M$ which are greater, but quite near
to $M_c$ (\ref{17}) (for example, at $M_c<M<1.01M_c$), there exists a
finite interval of the external magnetic field values, where stable
quark droplets are formed by massive quarks. But for values of the
external magnetic field outside of this interval, stable multiquark
droplets are composed of massless quarks.} The most interesting
results were obtained in the case $M=M_1$, where we have found that
an external magnetic field induces stable quark matter formation.
Indeed, at $H=0$ multiquark droplets are not stable in this case.
However, at $H\ne 0$ they are allowed to exist. Furthermore, there
exists a value of magnetic fields, $H_\alpha\sim 10^{19}$ Gauss, such
that at $H<H_\alpha$ quark droplets are formed by {\it massive}
quarks. At $H>H_\alpha$, multiquark droplets are composed of massless
quarks. We have also checked that in a similar way, for other values
of $M$ from the interval $M\in (\tilde M,M_c)$ (see Fig. 1), stable
quark droplet formation is also induced by the external magnetic
field.

Note that one can even assert that at $H\ne 0$ stable quark droplets
are allowed to exist in the initial NJL model at {\it arbitrary}
values of the coupling constant $G>0$. At $G>G_{bag}$ it is the
native property of this theory even at $H=0$, but at $G<G_{bag}$ the
stability of the quark matter is induced (catalized) by the external
magnetic field.

Moreover, using an energetical approach, we have shown that an
external magnetic field promotes the creation as well as enhances the
stability of quark droplets at $G>G_{bag}$.

Finally, let us add some remarks about the role of color
superconductivity (CSC) in the quark matter stability problem
considered above. This phenomenon was predicted quite recently in the
framework of instanton-induced- as well as NJL-type models with four
-- fermionic interactions (see the reviews \cite{nardulli} or some of
the recent papers \cite{csc,ebklim} and references therein).
At zero temperature it is proved that CSC must exist at
asymptotically large values of the chemical potential (or baryonic
densities), however, the lower boundary for this effect is not
established up to now. In particular, in \cite{sad} CSC was
investigated just in a slightly extended version of
the NJL model (1) at neighboring to the set 1 values of parameters,
i.e. for $M\approx M_1$ and $\Lambda =650$ MeV, but with an
additional four-quark term included, which is
responsible for the interactions in the diquark channel
(with a corresponding coupling constant $G^{'}$). It was
realized at $H=0$, that usually with growing value of $\mu$, the
chiral invariance of the system is restored before
or just at the CSC phase transition. Moreover, at fixed $G$, the
parameters of the CSC phase transition strongly depend on the value
of $\kappa\equiv G^{'}/G$, while the features of the chiral phase
transition do not seriously depend on $\kappa$. For example, in the
NJL model (1) at small values of $\kappa\le 1/4$ the transition to
the CSC phase occured at $\mu\sim 700$ MeV, but at $\kappa\sim 1$
it is at $\mu\sim 300$ MeV, whereas for wide interval of
$\kappa$-values ($G$ is fixed) the chiral
symmetry is restored at $\mu\sim 300$ MeV \cite{sad}.

For simplicity, we have put $G^{'}=0$ in our consideration of the
quark matter stability problem at $H\ne 0$. In the light of the above
discussion we believe that our results are not qualitatively changed
at least for small values of $\kappa$, i.e. in the case, where CSC
occurs for rather large values of $\mu\sim 700$ MeV \cite{sad}.
In fact, in this case CSC can not seriously influence stable quark
droplet formation which takes place at values of the chemical
potential much less than $700$ MeV (see the graphs of $\mu_c$ on Figs
3-5).  At present time there are several indications on the
possibility for CSC to take place only at chemical potential values
which indeed are outside the scope of our consideration, i.e. at $\mu
>> 300$ MeV: i) the CSC phenomenon was not discovered up to now in
heavy ion collision experiments, ii) It was argued in \cite{ar} that
two-flavor CSC is absent inside quark cores of neutron stars, where
baryonic density is several times the density $\rho_o$ of ordinary
nuclear matter, iii) In \cite{carter} it was shown that due to the
color Meissner effect as well as due to the presence of background
chromomagnetic fields (gluon condensate) the CSC is shifted to higher
densities (chemical potentials),
iv) As in the case with ordinary superconductivity, the sufficiently
strong external magnetic field prevents a system from CSC as well
etc. In these cases the ordinary, i.e. without CSC, quark matter
might be formed at rather low densities. To describe such a processes
one could use models of the type (1) with zero or very small value of
the diquark channel coupling constant $G^{'}$.

Recently, we have studied already the magnetic catalysis of the quark
droplet stabilization in the scope of a simpler one-flavor NJL model
\cite{ke}. Next, we are planning to consider the quark matter
stability problem in more realistic NJL models including vector
interactions as well as the third flavor of quarks etc.

\section*{Acknowledgments}

The authors are grateful to Dr. M.L. Nekrasov for help in numerical
calculations. This work was supported in part by DFG-project 436 RUS
113/477/0-2.

\newpage
\begin{center}
{\bf Figure Captions}
\end{center}
\vspace{0.5cm}

{\bf Fig. 1.} Schematic phase portrait of the NJL model at nonzero
$\mu$ and
for arbitrary values of the quark mass $M$ at $\mu =0$ (the case
$G>G_{crit}$).
Phases B and C are  
massive nonsymmetric phases, A is a chirally symmetric and massless
phase. In the phase B the quark mass is equal to $M$.
In the phase B the particle density in the ground state is equal to
zero, whereas in phases A, C it is nonzero.
 Solid and dashed lines represent critical curves of second and
first order phase transitions, respectively;
$\alpha$ and $\beta$ denote tricritical points. In particular,
the curve $l_{BC}$ has the following form:
$l_{BC}=\{(M,\mu):\mu=M\}$, the curve $l_{AB}$ is defined by the
equation (\ref{16}), $\tilde M=0.31\Lambda$,
$M_c=0.56\Lambda$. Only the values $\mu_c(M)$ (see (\ref{17})) of the
chemical potential, corresponding to points of the critical line
$l_{AB}$, obey the stability constraints (\ref{stab}).
\vspace{0.5cm}

{\bf Fig. 2.} Schematic $(H,\mu)$-phase portrait of the NJL model at
$M=M_1\equiv 0.48 \Lambda$ ($\tilde M<M_1<M_c$), where A is the
chirally symmetric massless phase, B, C and C$_1$ are chirally
noninvariant massive phases. Ground state baryonic density is zero in
the phase B, and nonzero in phases A, C, C$_1$. Dashed and solid
lines are critical curves of the first- and second order
phase transitions, respectively. ($l_{AB}$ is defined
by the equation (\ref{26})) The
curve L=$\{(H,\mu):\mu=m_o(H)\}$, where $m_o(H)\equiv
m_{vac}$ is the quark mass at $\mu=0$. $\mu_1\equiv 326$ MeV is such
a value of
the chemical potential that the point $(M_1,\mu_1)$ lies on the curve
$\overbrace{\alpha\beta}$ of Fig. 1. The external magnetic field
values, corresponding to the
tricritical points $\alpha, \beta, \gamma$ are the following:
$\sqrt{2|e|H_\alpha}/ \Lambda=0.85$, $\sqrt{2|e|H_\beta}/
\Lambda=0.82$ and $\sqrt{2|e|H_\gamma}/ \Lambda=0.67$. Only the
points of the critical lines $l_{AB}$ and $l_{BC}$ obey the stability
constraints (\ref{stab}).
\vspace{0.5cm}

{\bf Fig. 3.} The behaviour of quantities $m_{vac}\equiv m_o(H)$,
$\mu_c$ and $m_{dense}$ vs $H$ which
characterize the coexistence of the phase B with A and C ones in the
case $M=M_1$ and $\Lambda =650$ MeV.  If $\sqrt{2|e|H}/
\Lambda >0.85$ these parameters describe the first order phase
transition (curve $l_{AB}$) between A and B (in this case
$m_{dense}\equiv 0$ and $\mu_c\equiv \mu_c(H)$, where $\mu_c(H)$ is
the solution of the equation (\ref{26})), if $\sqrt{2|e|H}/ \Lambda
<0.85$ -- the phase transition between B and C (in this case
$m_{dense}\equiv m_1(H)$ and  $\mu_c\equiv \tilde\mu_c(H)$, where
$\tilde\mu_c(H)$ is defined by the equation (\ref{261})). Only the
part of $l_{BC}$, lying in the region $\omega_0$ is considered. These
values of $m_{vac}$, $\mu_c$ and $m_{dense}$ are the solutions of the
stability constraints (\ref{stab}).
\vspace{0.5cm}

{\bf Fig. 4.} The same as on Fig. 3, but for $M=M_2$ and $\Lambda
=600$ MeV (in this case $m_{dense}\equiv 0$). The dotted horizontal
line corresponds to the critical chemical potential value $\mu_c$ at
zero external magnetic field. For these values of $m_{vac}$, $\mu_c$
and $m_{dense}$ the stability constraints (\ref{stab}) are fulfilled.
\vspace{0.5cm}

{\bf Fig. 5.} The same as on Fig. 3, but for $M=M_3$ and $\Lambda
=570$ MeV (in this case $m_{dense}\equiv 0$ as well). The dotted
horizontal line corresponds to the critical chemical potential value
$\mu_c$ at zero external magnetic field. For these values of
$m_{vac}$, $\mu_c$ and $m_{dense}$ the stability constraints
(\ref{stab}) are fulfilled.
\vspace{0.5cm}

{\bf Fig. 6.} The energy per baryon in the NJL model (1) at $M=M_2$,
$\Lambda =600$ MeV for $H=0$ and $\sqrt{2|e|H}/ \Lambda=1.04$.
The solid line corresponds to the massive solutions of the
stationarity equation, the pointed line to the massless one.
\vspace{0.5cm}

{\bf Fig. 7.} The energy per baryon in the NJL model (1) at $M=M_1$,
$\Lambda =650$ MeV for $\sqrt{2|e|H}/ \Lambda=0.82$ and
$\sqrt{2|e|H}/ \Lambda=1.04$. 
The solid line corresponds to the massive solutions of the
stationarity equation, the pointed line to the massless one.
\vspace{0.5cm}

{\bf Fig. \ref{ensv}.} The binding energy per quark for a stable
droplet in the NJL model (1) vs external magnetic field in two cases:
$M=M_1$, $\Lambda=650$ MeV (solid line) and $M=M_2$, $\Lambda=600$
MeV (pointed line).
\vspace{0.5cm}

{\bf Fig. \ref{dens}.}
Baryon density $\rho_{drop}$ inside a stable
quark droplet in the NJL model (1) as a function of external magnetic
field. The solid line corresponds to the case $M=M_1$, $\Lambda=650$
MeV, the pointed one to the case $M=M_2$, $\Lambda=600$ MeV.
\vspace{0.5cm}

\begin{figure}
\includegraphics{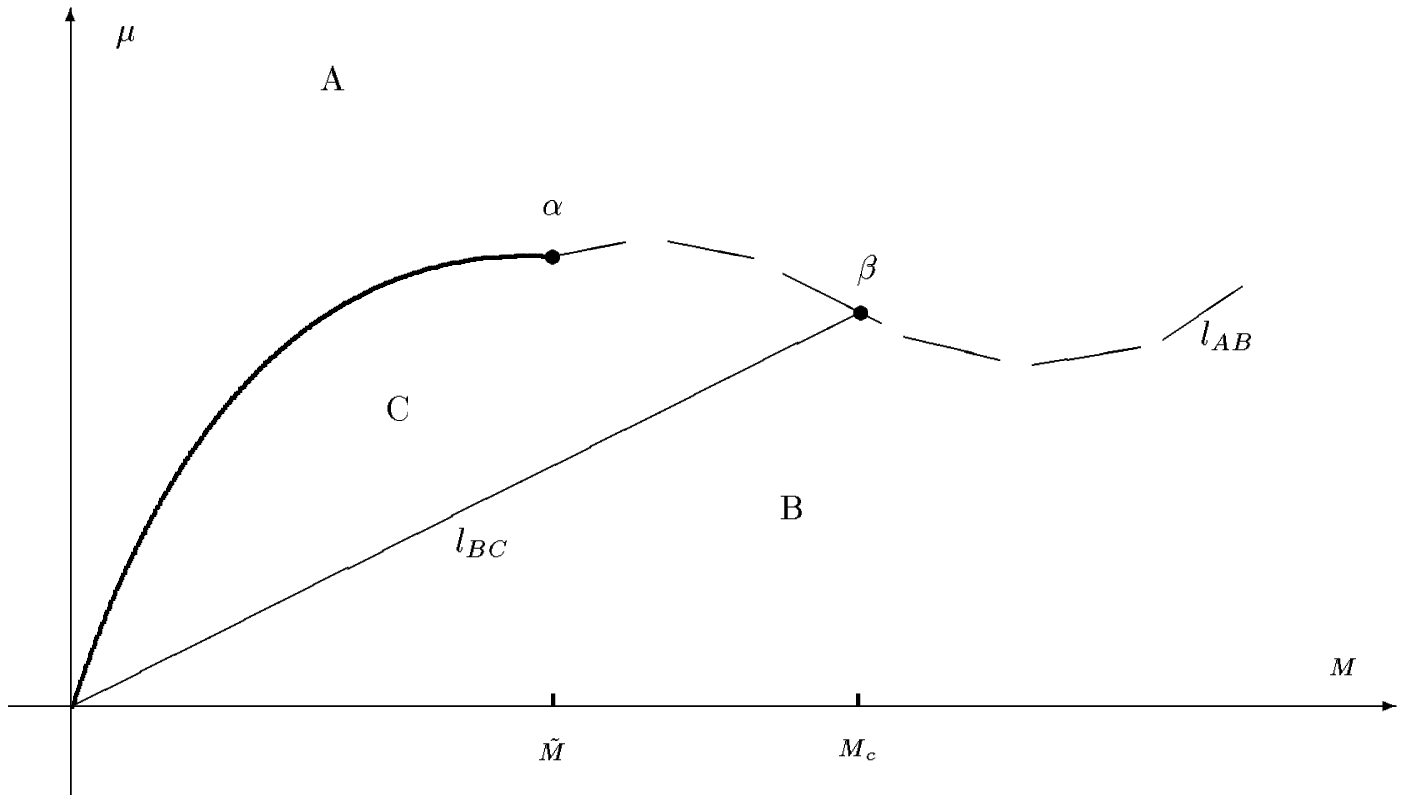}
\caption{}
\end{figure}

\begin{figure}
\includegraphics{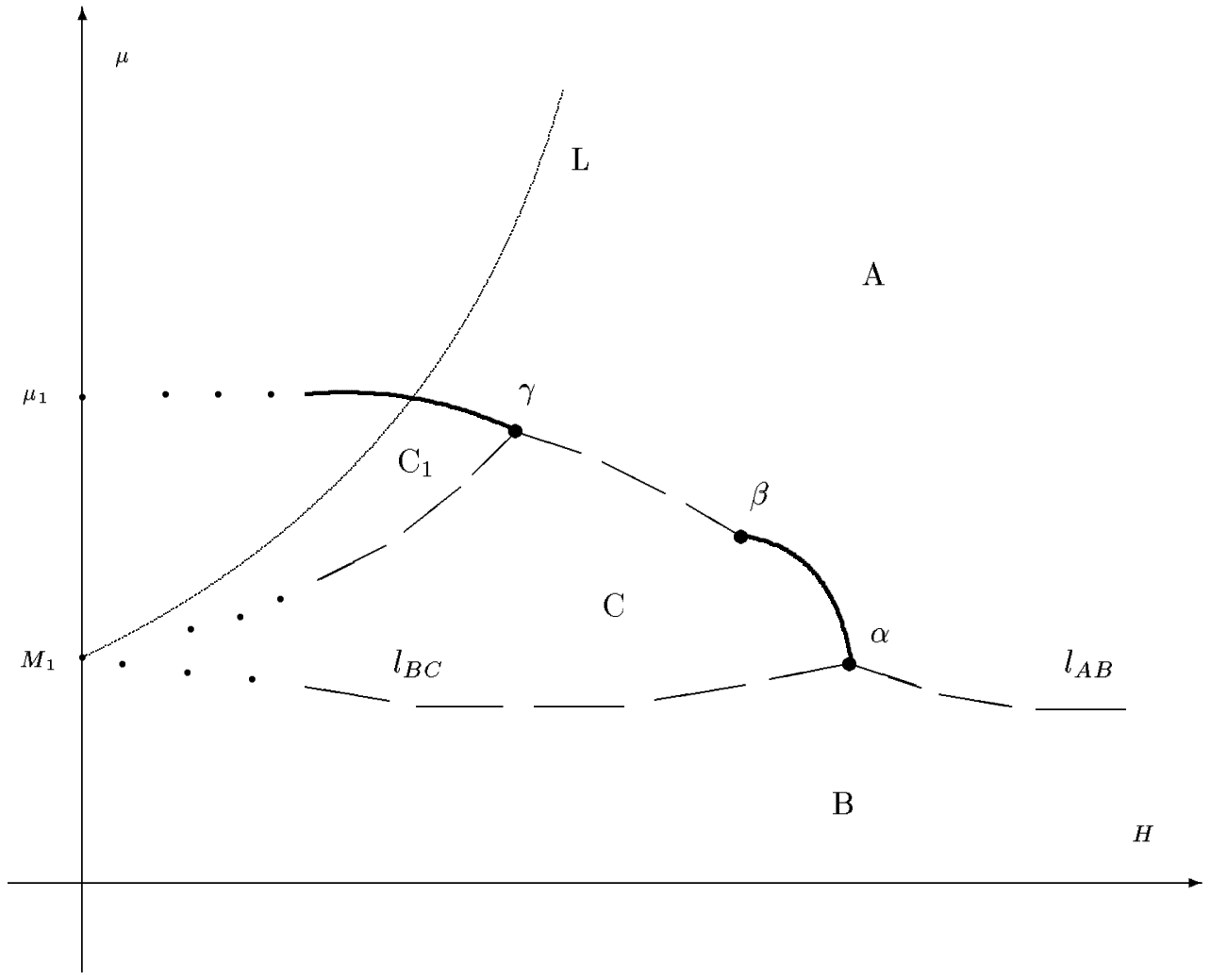}
\caption{}
\end{figure}

\begin{figure}
\includegraphics{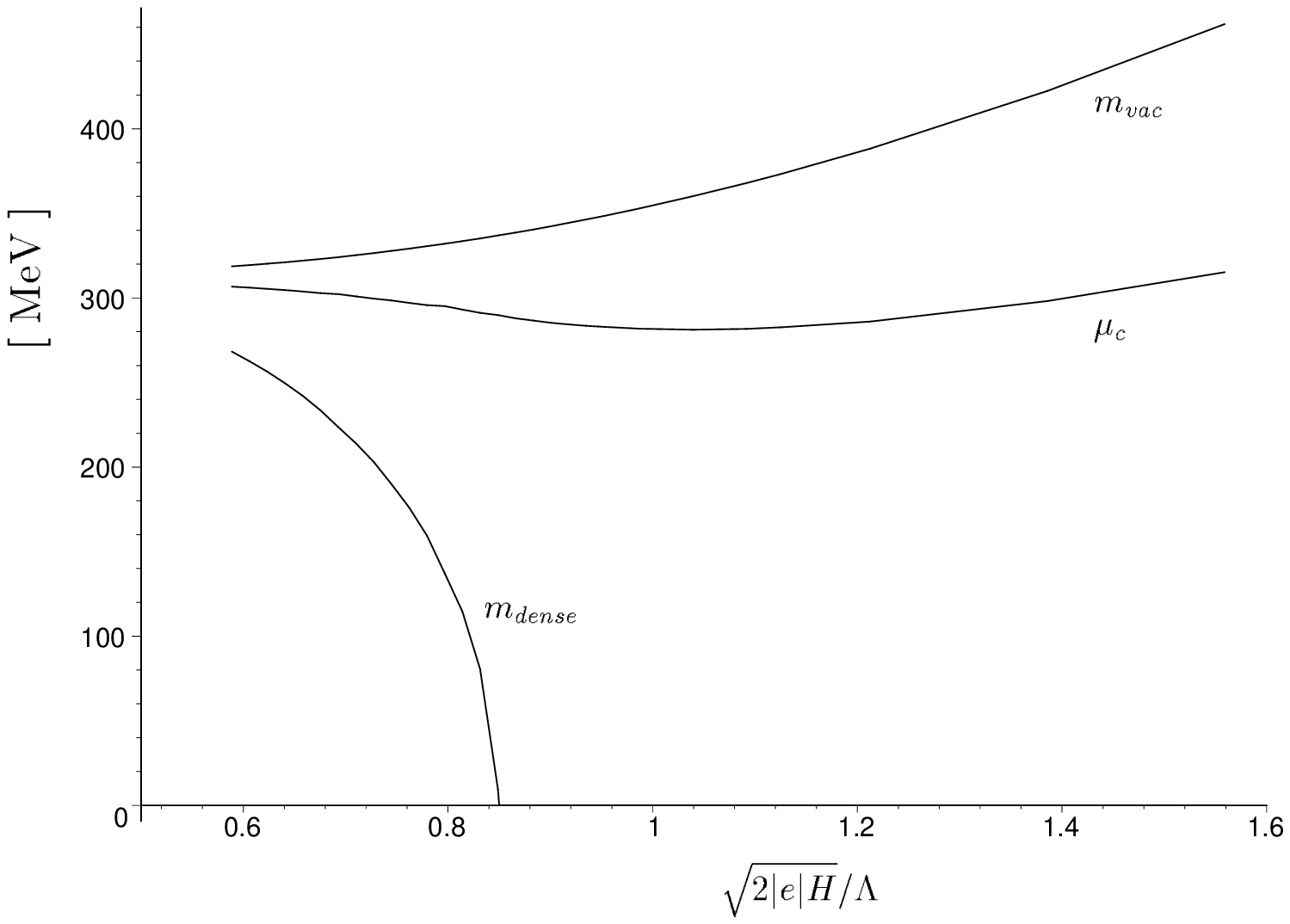}
\caption{}
\end{figure}

\begin{figure}
\includegraphics{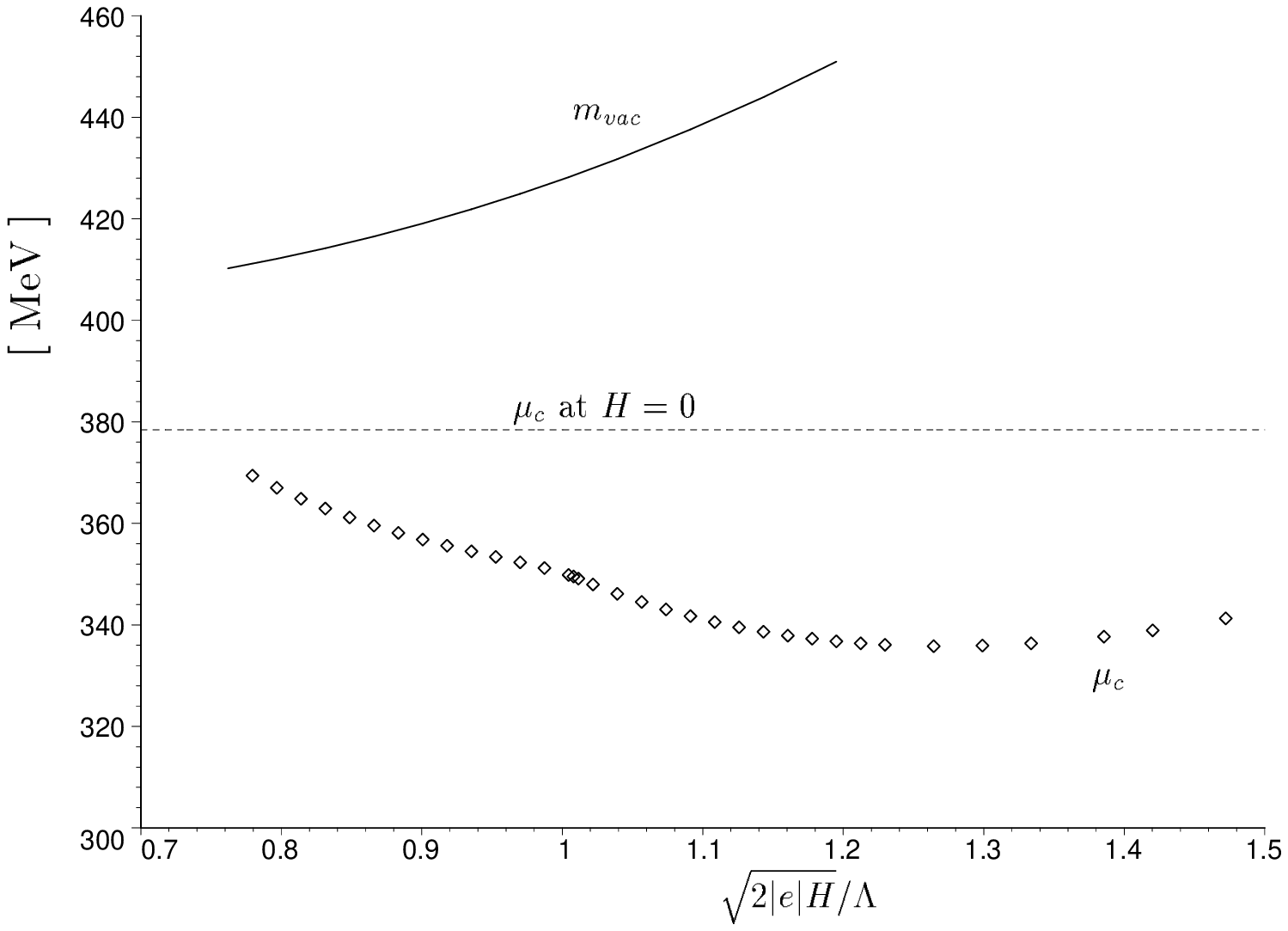}
\caption{}
\end{figure}

\begin{figure}
\includegraphics{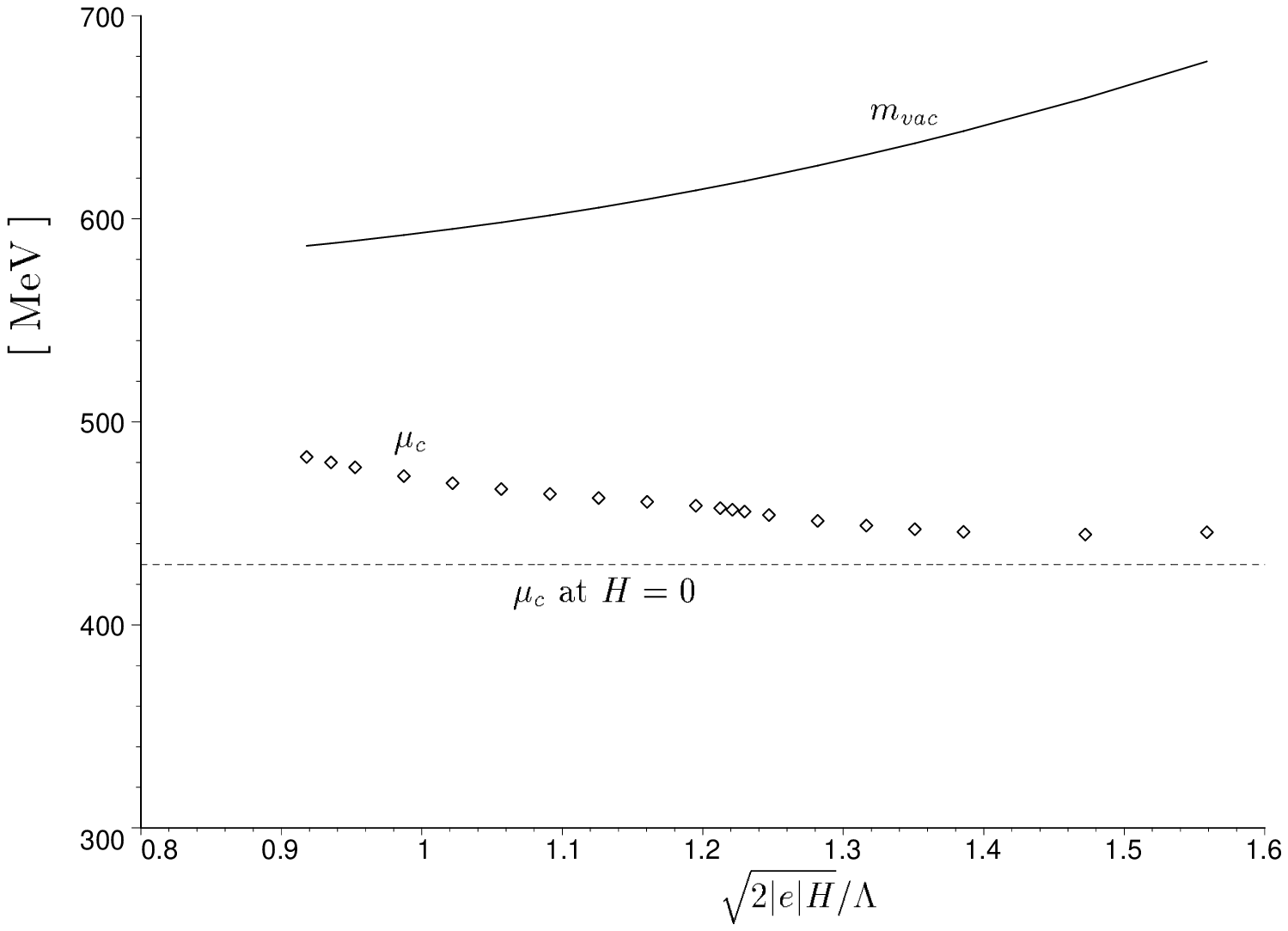}
\caption{}
\end{figure}

\begin{figure}
\includegraphics{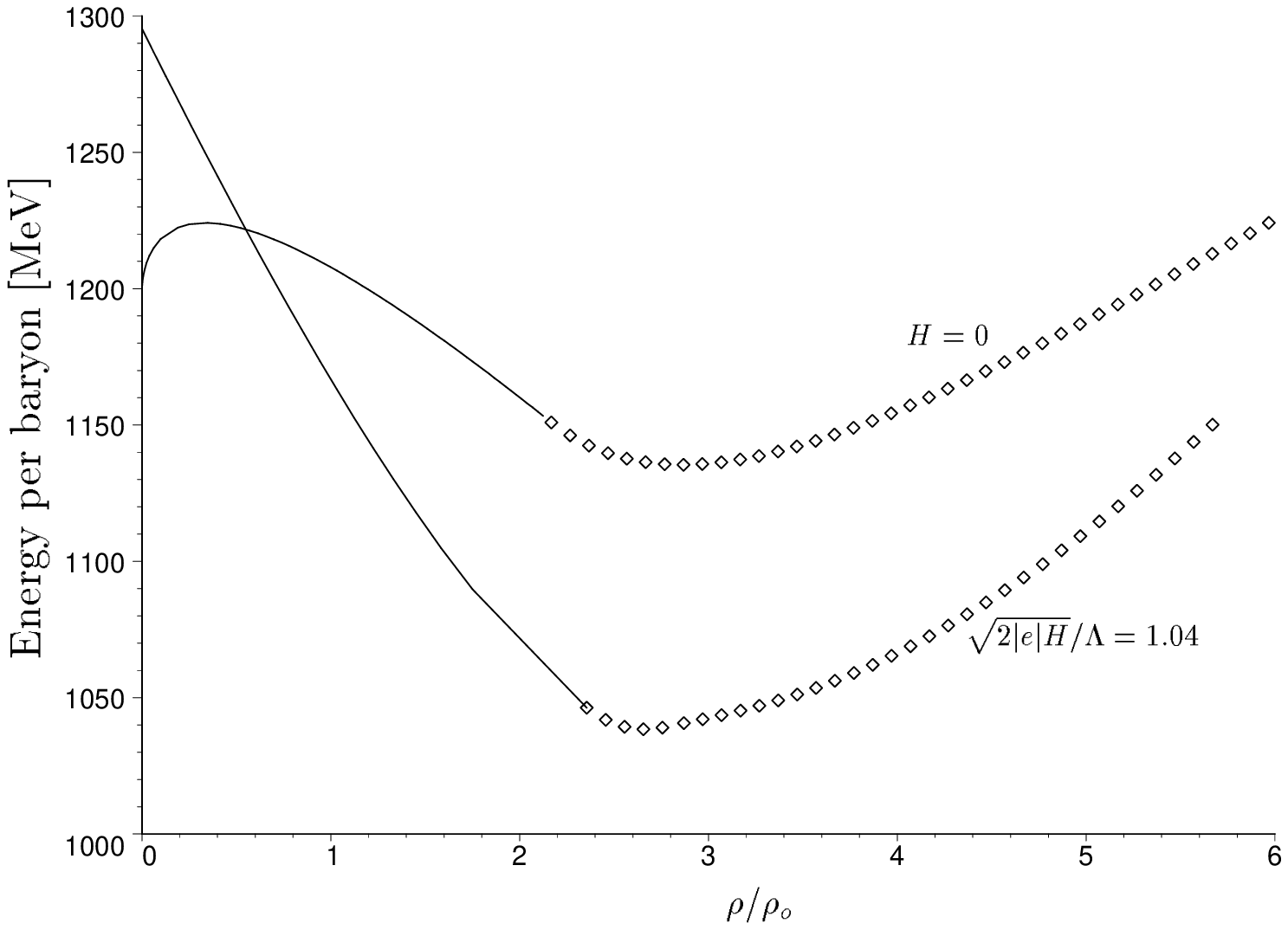}
\caption{}
\end{figure}

\begin{figure}
\includegraphics{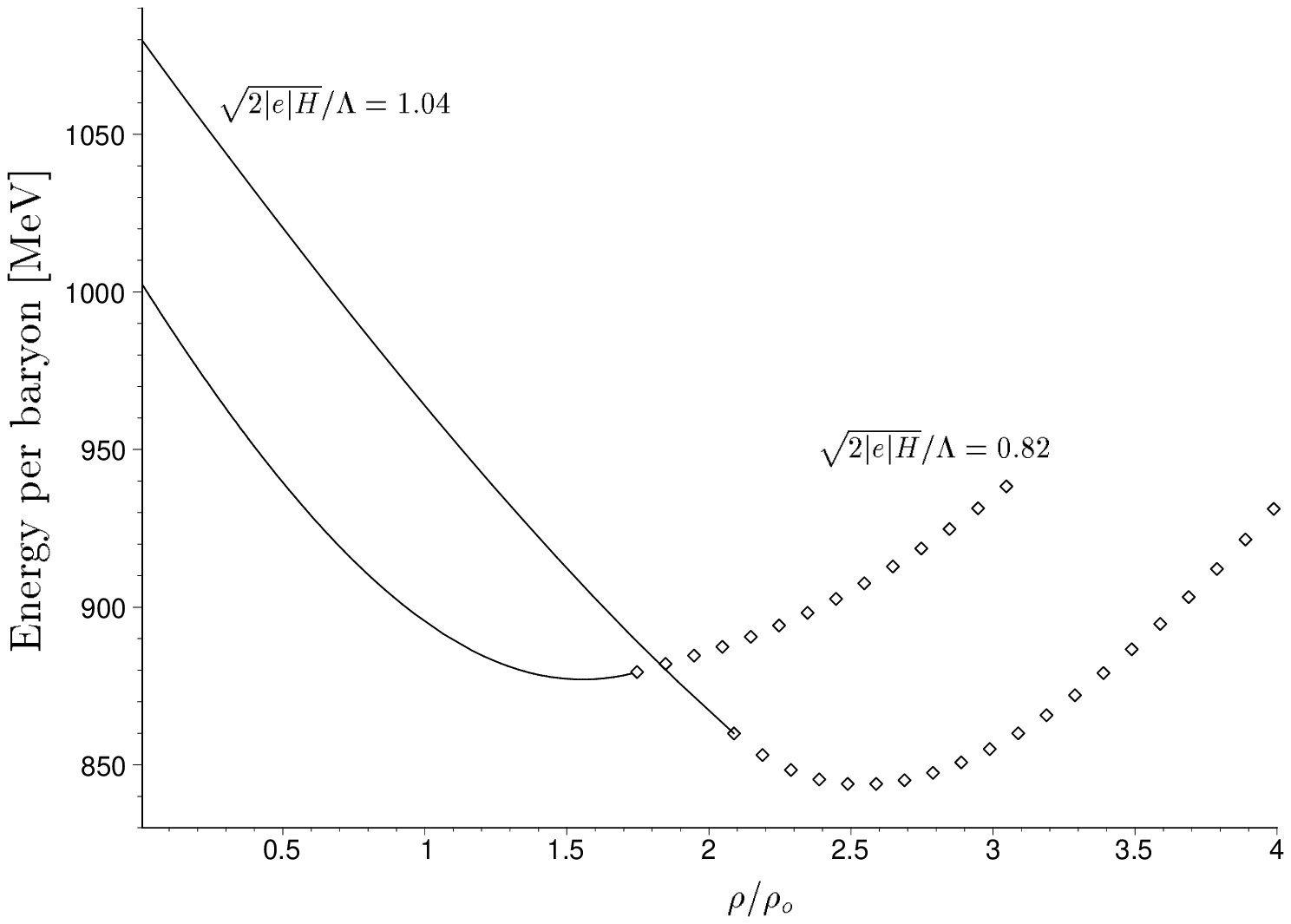}
\caption{}
\end{figure}

\begin{figure}
\includegraphics{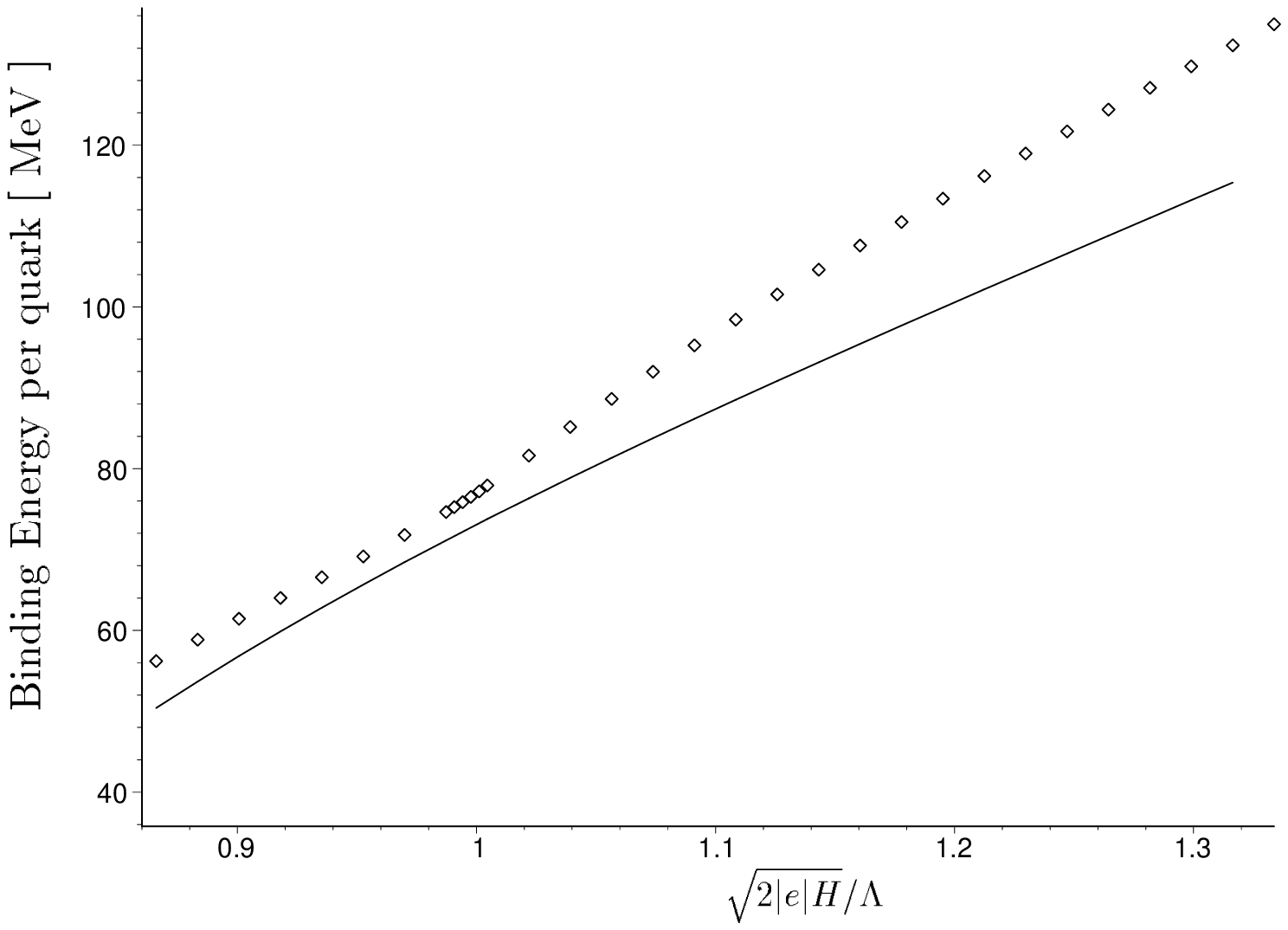}
\caption{\label{ensv}}
\end{figure}

\begin{figure}
\includegraphics{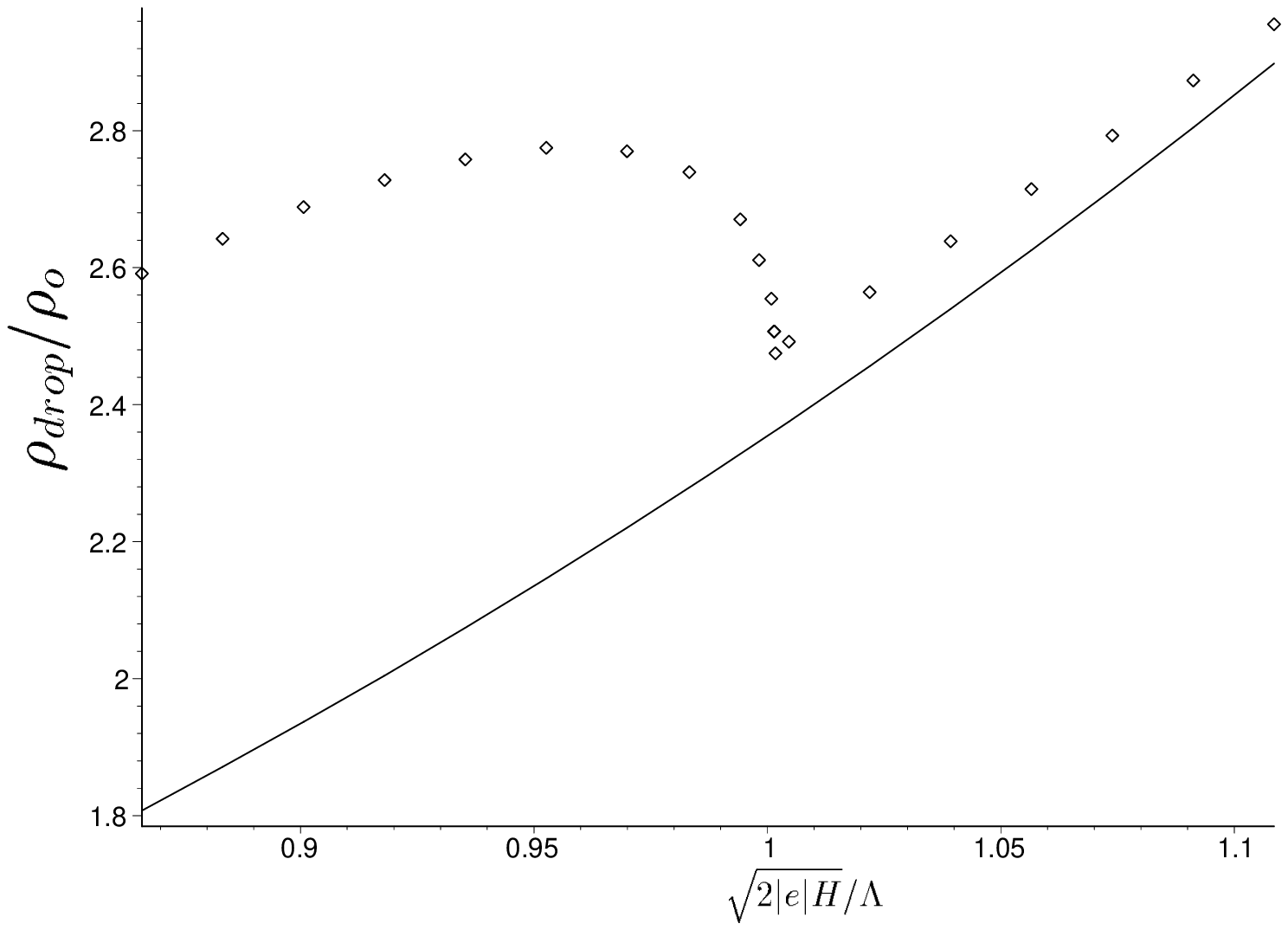}
\caption{\label{dens}}
\end{figure}


\begin{thebibliography}{99}

\bibitem{fodor}
Z. Fodor and S.D. Katz, Phys. Lett. {\bf B534} (2002) 87;
S. Ejiri, C.R. Allton, S.J. Hands et al,
hep-lat/0209012; M. D'Elia and M.P. Lombardo, Phys. Rev. {\bf D67}
 (2003) 014505; F. Karsch and E. Laermann, hep-lat/0305025.

\bibitem{wal}
J.D. Walecka, Ann. Phys. {\bf 83} (1974) 491.

\bibitem{kap}
J.I. Kapusta, {\it Finite-temperature field theory},
(Cambridge University Press,1989).

\bibitem{njl}
Y. Nambu and G. Jona-Lasinio, Phys. Rev. {\bf 122} (1961) 345.

\bibitem{ebvol}
D. Ebert and M.K. Volkov, Yad. Fiz. {\bf 36} (1982) 1265; Z.Phys.
{\bf C16} (1983) 205; D. Ebert and H. Reinhardt, Nucl. Phys. {\bf
B271} (1986) 188; D. Ebert, H. Reinhardt and M.K.Volkov, Progr.
Part. Nucl. Phys. {33} (1994) 1.

\bibitem{ebka}
D. Ebert, L. Kaschluhn and G. Kastelewicz, Phys. Lett. {\bf B264}
 (1991) 420.

\bibitem{vog}
U. Vogl, Z. Phys. {\bf A337} (1990) 191; U. Vogl and W. Weise,
Progr. Part. Nucl. Phys. {\bf 27} (1991) 195.

\bibitem{ebjur}
D. Ebert and L. Kaschluhn, Phys. Lett. {\bf B297} (1992) 367;
D. Ebert and T. Jurke, Phys. Rev. {\bf D58} (1998) 034001;
L.J. Abu-Raddad et al., Phys. Rev. {\bf C66} (2002) 025206.

\bibitem{reinh}
H. Reinhardt, Phys. Lett. {\bf B244} (1990) 316;
N. Ishii, W. Benz and K. Yazaki, Phys. Lett. {\bf B301} (1993) 165;
C. Hanhart and S. Krewald, ibid. {\bf 344} (1995) 55.

\bibitem{hk}
T. Hatsuda and T. Kunihiro, Phys. Rept. {\bf 247} (1994) 221.

\bibitem{ratti}
W.M. Alberico, F. Giacosa, M. Nardi and C. Ratti, Eur. Phys. J. {\bf
A16} (2003) 221.

\bibitem{oettel}
M. Oettel, G. Hellstern, R. Alkhofer and H. Reinhardt,
 Phys. Rev. {\bf C58} (1998) 2459.

\bibitem{bub}
M. Buballa, Nucl. Phys. {\bf A611} (1996) 393; M. Buballa and M.
Oertel, Nucl. Phys. {\bf A642} (1998) 39.

\bibitem{mark}
M. Alford, K. Rajagopal and F. Wilczek, Phys. Lett. {\bf B422}
 (1998) 247.

\bibitem{satarov}
M. Buballa and M. Oertel, Phys. Lett. {\bf B457} (1999) 261;
I.N. Mishustin, L.M. Satarov, H. Stucker et al,
Phys. Atom. Nucl. {\bf 64} (2001) 802;
C. Ratti, Europhys. Lett. {\bf 61} (2003) 314.

\bibitem{chodos}
A. Chodos, R.L. Jaffe, K. Johnson and C.B. Thorn, Phys. Rev. {\bf
D10} (1974) 2599.

\bibitem{chak}
A.S. Vshivtsev and D.V. Serebryakova, Zh. Eksp. Teor. Fiz. {\bf 106}
 (1994) 35; G.G. Likhachev and A.I. Studenikin, JETP {\bf 81}
(1995) 419; S. Chakrabarty, Phys. Rev. {\bf D54} (1996) 1306;
T.C. Phukon,  Phys. Rev. {\bf D62} (2000) 023002;
V.R. Khalilov, Phys. Rev. {\bf D65} (2002) 056001; Teor. Mat. Fiz.
{\bf 133} (2002) 103.

\bibitem{tatsumi}
T. Tatsumi, Phys. Lett. {\bf B489} (2000) 280; astro-ph/0004062.

\bibitem{berg}
M. Alford, J. Berges and K. Rajagopal, Nucl. Phys. {\bf B571}
(2000) 269.

\bibitem{blashke}
D. Blaschke, D.M. Sedrakian and K.M. Shahabasyan et al, Astrofiz.
{\bf 44} (2001) 443.

\bibitem{k3}
K.G.~Klimenko, preprint IHEP 98-56, Protvino 1998 [hep-ph/9809218];
A.S. Vshivtsev, M.A.~Vdovichenko and K.G.~Klimenko,
Phys. Atom. Nucl. {\bf 63} (2000) 470.

\bibitem{k2}
 D.~Ebert et al., Phys. Rev. {\bf D61} (2000) 025005; D. Ebert,
 M.A.~Vdovichenko
and K.G.~Klimenko, Phys. Atom.  Nucl. {\bf 64} (2001) 336; D. Ebert
and K.G.~Klimenko, nucl-th/9911073.

\bibitem{kiriyama}
J. Madsen, Phys. Rev. {\bf D47} (1993) 5156; E.P. Gilson and R.L.
Jaffe, Phys. Rev. Lett. {\bf 71} (1993) 332;
O. Kiriyama and A. Hosaka, Phys. Rev. {\bf D67} (2003) 085010.

\bibitem{zhuang}
P. Zhuang, J. Hufner and S.P. Klevansky, Nucl. Phys. {\bf A576}
(1994) 525.

\bibitem{k1}
A.S. Vshivtsev and K.G. Klimenko, JETP Lett. {\bf 64} (1996) 338
[hep-ph/9701288];
A.S. Vshivtsev, V.Ch. Zhukovsky and K.G. Klimenko, JETP {\bf
84} (1997) 1047.

\bibitem{bateman}
H. Bateman and A. Erdeyi, {\it Higher Transcendental Functions,}
(McGrawHill, New York, 1953).

\bibitem{mir}
K.G. Klimenko, Theor. Math. Phys. {\bf 89} (1991) 1161; {\bf 90}
(1992) 1; Z. Phys. {\bf C54} (1992) 323; I.V. Krive and S.A.
Naftulin,
Phys. Rev. {\bf D46} (1992) 2737; V.P. Gusynin, V.A. Miransky and
I.A. Shovkovy, Phys. Rev. Lett. {\bf 73} (1994) 3499; Phys. Lett.
{\bf B349} (1995) 477.

\bibitem{gus}
V.A. Miransky, Progr. Theor. Phys. Suppl. {\bf 123} (1996) 49;
A.S. Vshivtsev et al, Phys. Part. Nucl. {\bf 29} (1998) 523;
V.P. Gusynin, Ukr. J. Phys. {\bf 45} (2000) 603;
V. de la Incera, hep-ph/0009303. 


\bibitem{incera}
V.C. Zhukovsky et al, JETP Lett. {\bf 73} (2001) 121;
Theor. Math. Phys. {\bf 134} (2003) 254;
E.J. Ferrer, V.P. Gusynin and V. de la Incera, cond-mat/0203217; V.A.
Miransky, hep-ph/0208180;
E. Elizalde, E.J. Ferrer and V. de la Incera, hep-ph/0209324;
V.A. Miransky and I.A. Shovkovy, Phys. Rev. {\bf D66} (2002) 045006;
V.P. Gusynin, V.A. Miransky and I.A. Shovkovy, hep-ph/0304059.

\bibitem{duncan}
R.C. Duncan, astro-ph/0002442; S. Ghosh et al, astro-ph/0106153.

\bibitem{nardulli}
M. Alford, Ann. Rev. Nucl. Part. Sci. {\bf 51} (2001) 131;
B.O. Kerbikov, hep-ph/0110197; G. Nardulli,  Riv. Nuovo Cim. {\bf
25N3} (2002) 1.

\bibitem{csc}
S.M. Molodtsov and G.M. Zinovjev, Mod. Phys. Lett. {\bf A18}
(2003) 817; hep-ph/0112075;
B.O. Kerbikov, Phys. Atom. Nucl. {\bf 65} (2002) 1918;
M. Huang, P. Zhuang and W. Chao, Phys. Rev. {\bf D65} (2002) 076012;
hep-ph/0110046; E. Nakano, T. Maruyama and T. Tatsumi, 
hep-ph/0304223.

\bibitem{ebklim}
D. Ebert, K.Klimenko and H.Toki, Phys.Rev. {\bf D64}
(2001) 014038; D. Ebert et al. Prog. Theor. Phys. {\bf 106} (2001)
835; JETP Letters {\bf 74} (2001) 523.

\bibitem{sad}
T.M. Schwarz, S.P. Klevansky and G. Papp, Phys. Rev. {\bf C60}
 (1999) 055205; M. Sadzikowski, Mod. Phys. Lett. {\bf A16}
(2001) 1129.

\bibitem{ar}
M. Alford and K. Rajagopal, hep-ph/0204001.

\bibitem{carter}
N.O. Agasian, B.O. Kerbikov and V.I. Shevchenko, Phys. Rept. {\bf
320} (1999) 131;
G.M. Carter and D. Diakonov, Nucl. Phys. {\bf B582} (2000) 571;
 D. Ebert et al, Phys. Rev. {\bf D65} (2002) 054024.

\bibitem{ke}
K.G. Klimenko and D. Ebert, {\it Magnetic catalysis of the quark
matter stability in the Nambu -- Jona-Lasinio model}, preprint IFVE
2003-5, Protvino (2003) [in Russian].

\end{thebibliography}
\end{document}